\documentclass[a4paper,12pt]{article}
\usepackage{setspace}

\usepackage[centertags]{amsmath}
\usepackage{amssymb}
\usepackage{cmap}
\usepackage[utf8]{inputenc}
\usepackage[T2A]{fontenc}

\usepackage{graphicx}
\usepackage{epsfig}
\usepackage[english]{babel}
\usepackage{array}
\usepackage{amsthm}
\usepackage{latexsym}

\usepackage{mathrsfs}
\usepackage[mathcal]{euscript}

\pdfoutput=1
\usepackage{epsfig}
\usepackage{multirow}

\newcommand{\be}{\begin{equation}}
\newcommand{\ee}{\end{equation}}

\begin{document}

\title{On the theory of relativistic Brownian motion}
\vspace{1cm}
\author{
E. A. Kurianovich, A.I. Mikhailov, I. V. Volovich}
\date {~}
\maketitle
\begin{center}
  \small{\it Steklov Mathematical Institute of Russian Academy of Sciences,
st. Gubkina 8, 119991, Moscow, Russia.\\ Russian Society for the History
and Philosophy of Science.\\
1/36 Lyalin lane, bd. 2,
Moscow 105062,
Russian Federation.\\
 Philosophy department of Lomonosov Moscow State University, Leninskie Gory, Moscow, 119991, Russian
Federation.}\\
\small{\it email:\texttt{kurianovich@mail.ru,\ mikhailov1984@gmail.com,\ volovich@mi-ras.ru}}
\end{center}

\abstract{The approach to the theory of a relativistic random process is considered
    by the path integral method as Brownian motion taking into account the boundedness of
  speed. An attempt was made to build
   a relativistic analogue of the Wiener measure as a weak limit of finite-difference approximations. A formula has been proposed for calculating the probability
particle transition during relativistic Brownian motion. Calculations were carried out by three different methods with identical results. Along the way, exact and asymptotic formulas for the volume of some parts and sections of an N-1-dimensional unit cube were obtained. They can
have independent value.}


\doublespacing






\newtheorem{lemma}{Lemma}[section]
 \newtheorem{theorem}{Theorem}[section]


\section{Introduction}

The mathematical theory of non-relativistic Brownian motion was
developed by Wiener and others and is one of
the most important parts of the theory of random processes \cite{NV, BS, Strat, Ni, TH}. P-adic Brownian motion is considered in  section 16.2 of book \cite{V} and in articles \cite{D, Bikulov_Volovich}, see also references therein. Path integrals in p-adic space have been studied in \cite{Vladimirov_Volovich, Smolyanov_Shamarov, Shamarov}.
The concept of speed in Brownian motion is considered in \cite{AG, Ni, TH, LA}.
Theory of
relativistic Brownian motion is still in
early stage of development, despite numerous publications
on this issue, see \cite{DH, AS} for review.

In this paper, we consider finite-difference approximations
for the formal path integral  \cite{fe} defining the analogue
Wiener measures in the relativistic case. We consider the question:
is it possible to give a mathematical meaning to the expression \eqref{7777}?
\begin{eqnarray}
I(t_2,x_2;t_1,x_1) =\int \exp (-S[t,x,\dot{x}])Dx \label{7777},
\end{eqnarray}
where $t_1 , t_2 \in \mathbb{R} : t_2 > t_1 >0 $; $x_1 , x_2 \in \mathbb{R} $. If  $S$ is the action for a non-relativistic particle on a line:
\be
S_{nr}[t,x,\dot{x}]=\int_{t_1}^{t_2}\frac{m\dot{x}^2}{2} dt,\label{}
\ee
where $x: [t_{1}; t_{2}] \to \mathbb{R},$ $\dot{x}=\frac{dx}{dt}$, $m$ is the mass of the Brownian particle. Eq \eqref{7777} can be interpreted mathematically the density of transition probability from the point $x_{1}$ at time $t_{1}$ to the point $x_{2}$ at time $t_{2}$ as it was proved by Wiener \cite{NV}. Action for relativistic Brownian motion \cite{L} is
\begin{eqnarray}
S[t,x,\dot{x}]=-mc^{2}\int_{t_1}^{t_2}\sqrt{1-\frac{\dot{x}^2}{c^{2}}} dt\label{}.
\end{eqnarray}
Here  $c$ is the speed of light.

The task is to study the possibility of describing the transition
probabilities of relativistic Brownian motion using
functional integral. In Section 2 of the article, the value of the regularizing factor was selected using the polyhedron method. Along the way, exact and asymptotic formulas are derived for the volume of some parts and sections of an N-1-dimensional unit cube, which can have independent value. In Section 3, the value of the regularizing factor and the asymptotic formula for the desired sequence of path integrals are obtained up to a pre-exponential factor. The calculation is carried out using the recurrent convolution method. Section 4 performs direct and inverse Fourier transforms of the sequence of integrals. The saddle point method is used to estimate the inverse Fourier transform. By this method is obtained the asymptotic value of the sequence of integrals, including the pre-exponential factor. The regularizing factor and the value of the regularized limit are also obtained. A formula is proposed for calculating the transition probability of a particle during Brownian motion. It is shown that the results obtained by three different methods coincide.

\section{Method of polyhedra}
Let us consider one-dimensional Brownian motion when a particle moves from point $a_1$ to point $a_2$ in time $ t_{2}-t_{1}$. Let's divide the time interval into N parts and write N-fold approximations for the path integral:
\begin{eqnarray}
b_{N}=b_{N}(a_2,t_2;a_1,t_1)=C_{N}
\int_{G_{N}}\exp\{mc^{2}\varepsilon\sum_{j=0}^{N-1}
\sqrt{1-\frac{(x_{j+1}-x_{j})^{2}}{c^{2}\varepsilon^{2}}}\}
dx_{1}\ldots dx_{N-1}, \label{24}\end{eqnarray}
where
$$\varepsilon =\frac{t_{2}-t_{1}}{N}, t_{2}>t_{1},$$
$$ x_{0}=a_{1}, x_{N}=a_{2},$$
$$ G_{N}=\{(x_{1},\ldots, x_{N-1})\epsilon \mathbb{R} ^{N-1}:|x_{j+1}-x_{j}|\leq\varepsilon c,j=0,1,\ldots, N-1\}. $$
In addition, the following should be hold:
\be |\frac{a_{2}-a_{1}}{c(t_{2}-t_{1})}| \leq1.\ee
It is required to determine a value of $C_{N}$ such that the next limit of the sequence $b_{N}$ would be a finite positive value.
\be L=\lim_{N\rightarrow\infty}b_{N}=\lim_{N\rightarrow\infty}
C_{N}\int_{G_{N}}\exp\{mc^{2}
\varepsilon\sum_{j=0}^{N-1}
\sqrt{1-\frac{(x_{j+1}-x_{j})^{2}}{c^{2}\varepsilon^{2}}}\}
dx_{1}dx_{2}\ldots dx_{N-1}. \ee
It will be shown below that such $C_{N}=C_{N}(\beta)$ exists. Here
  \be0\leq\beta=\frac{1}{2}-\frac{a_{2}-a_{1}}{2c(t_{2}-t_{1})}\leq1.\label{135}\ee
Let's make the following change of variables:
$$ y_{j}=\frac{1}{2}-\frac{x_{j}-x_{j-1}}{2\varepsilon c}, j=1,...,N-1. $$
Then
$$b_N=\int_{G'_N}\exp\{2 m c^{2}\varepsilon \sum_{j=1}^{N-1}
\sqrt{y_{j}(1-y_{j})}+mc^{2}\varepsilon\sqrt{1-(\frac{a_{2}-a_{1}}{\varepsilon c}-N +1+2\sum_{j=1}^{N-1}y_{j})^{2}}\}\times$$

\be \times C_{N}(2\varepsilon c )^{N-1}dy_{1}dy_{2}\ldots dy_{N-1}. \label{25}\ee

Here
\be G'_{N}=\{(y_{1},\ldots, y_{N-1})\epsilon \mathbb{R}^{N-1}:0\leq y_{j}\leq1, \beta N-1\leq\sum_{j=1}^{N-1}y_{j}\leq \beta N .\label{200} \ee
Region $G'_{N}$ is a part of an N-1-dimensional unit cube, enclosed between two hyperplanes perpendicular to one of the diagonals of the cube. Let us take the point of intersection of this diagonal with the plane $\sum_{j = 1}^{N-1}y_{j}=\beta N$ and the point of intersection of the same diagonal with the plane $\sum_{j = 1}^{N- 1}y_{j}=\beta N-1.$ The first point has coordinates $$(\frac{\beta N}{N-1},\frac{\beta N}{N-1},... ,\frac{\beta N}{N-1}).$$ Second: $$(\frac{\beta N-1}{N-1},\frac{\beta N-1}{N-1 },...,\frac{\beta N-1}{N-1}).$$ The distance between two planes (equal to the distance between two points) will be: $$\sqrt{(\frac{\beta N} {N-1}-\frac{\beta N-1}{N-1})^{2}(N-1)}=\frac{1}{\sqrt{N-1}}.$$
   The length of a segment parallel to one of the edges of the cube with ends on these girder planes is 1.
\begin{theorem}
Let $t_{2}>t_{1}>0,\ 0\leq \beta \leq 1$,
\be C_{N}=\frac{A}{(2\varepsilon c )^{N-1}V_{N-1}}, \label{28} \ee
where $V_{N-1}$ is the volume of the region $G'_{\varepsilon}$, $A$ is a positive constant. Then the $L=\lim_{N\rightarrow\infty}b_{N}$ (limit of the sequence $b_{N}(\beta,\ t_{2}-t_{1})\ $ from \eqref{25} \  at\  $N\rightarrow\infty$) exists under these conditions and has the following estimate:
\be A\leq L\leq A e^{mc\sqrt{c^{2}(t_{2}-t_{1})^{2}-(a_{2}-a_{1})^{ 2}}}.\label{29}\ee
\end{theorem}
{\bf Proof.}
We examine the function in the exponent \eqref{25} for the smallest and largest value in the region $G'_{N}$. To begin with, we will neglect the last term tending to $0$. After this, it becomes clear that the function reaches its minimum value, equal to $0$, at any vertex of the cube lying in the region $G'_{N}$ ($y_{j}=0$ or $1$, $j=0 ,1, ... N-1 $). Now we examine the function for a conditional extremum in the plane
  \be\sum_{j=1}^{N-1}y_{j}= \beta N-\theta, 0\leq\theta\leq1.\label{27}\ee
Let's compose the Lagrange function:
$$F(y_{j})=2mc^{2}\varepsilon\sum_{j=1}^{N-1}\sqrt {y_{j}(1-y_{j})}+\lambda( \sum_{j=1}^{N-1}y_{j}- \beta N+\theta).$$
Let us equate all partial derivatives to $0$:
\be\frac{\partial F}{\partial y_{j}}=\frac{mc^{2}\varepsilon(1-2y_{j})}{\sqrt{y_{j}(1-y_{ j})}}+\lambda=0.\label{26}\ee
From \eqref{26} it follows that at the critical point all coordinates are equal, and, therefore, it is located at the intersection of the plane \eqref{27} and the diagonal of the cube to which this plane is perpendicular. From \eqref{27} we obtain the coordinates of the critical point for any $j$:
$$y_{j}=\frac{\beta N-\theta}{N-1}=\beta+O(\frac{1}{N}), N\rightarrow\infty.$$
All second unmixed derivatives (and mixed ones are equal to $0$) at the critical point are obviously negative. So this is the maximum point.
Substituting the values of the coordinates at the points of maximum and minimum values and \eqref{28} into \eqref{25} and using the mean value theorem, we get
\eqref{29}.$\Box$

The constant $A$ can be determined from the normalization:
$$ \int_{a_{1}-c(t_{2}-t_{1})}^{a_{1}+c(t_{2}-t_{1})}Lda_{2}=1. $$

The following theorem gives us the exact value of the volume $V_{N-1}$ of the region $G'_{N}$.
\begin{theorem}
  The formula for the volume of the region $G'_{N}$ from \eqref{200} is given by the following expression:
  \be V_{N-1}(\beta)=\sum_{k=0}^{[\beta N]}\frac{(-1)^{k}(\beta N-k)^{N-1} N}{k!(N-k)!},\label{30} \ee
  where $\beta $ is from \eqref{135}. $[\beta N]$ is the integer part of the number $\beta N$.
\end{theorem}
{\bf Proof.}

Volume of an arbitrary convex polyhedron in N-dimensional space: $$V=\frac{1}{N}\sum h_{i}S_{i},$$ where $h_{i}$ is the distance from an arbitrary point inside or on surface of the polyhedron to its $i$ -th face, $S_{i}$ is the area of this face, summation is carried out over all faces.

Let the plane $\sum_{i=1}^{N}x_{i}=d$ intersect the diagonal of an N-dimensional cube\\ $0\leq x_{i}\leq1$ (for example, a 3-dimensional cube $ ABCDA_{1 }B_{1}C_{1}D_{1}$) at the point O and cuts off the convex polyhedron $V(N,d),$ for which $\sum_{i=1}^{N}x_{i}\leq d$. Let's divide all the faces of the cube into two groups: the first group - in the planes $x_{i}=0$ (for $N=3$ these are the faces $ABCD, ABB_{1}A_{1}, ACC_{1}A_{1} ,$ containing point A), the second group - faces in planes $x_{i}=1,$ (containing point $D_{1}$ ). Then the distance from the point O to the faces of the first type is $\frac{d}{N}$ , and to the faces of the second type $1-\frac{d}{N}$ . Let $S_{0}$ be the area of that part of the face $x_{N}=0$ (face $ABCD$) that belongs to $V(N,d),$ and similarly $S_{1}$ be the area of intersection of the face $x_{N}=1$ (faces $A_{1}B_{1}C_{1}D_{1}$ ) and $V(N,d),$ . Note that the same area $S_{0},$ is cut off from all faces of the first group and the same for the second group. Because faces in each group of $N$ pieces, we get:
$$V(N,d)=\frac{1}{N}(N\frac{d}{N}S_{0}+N(1-\frac{d}{N})S_{1}) =\frac{dS_{0}+(N-d)S_{1}}{N}.$$

Now let's find $S_{0}$. Note that the face $x_{N}=0$ is an N-1-dimensional “hypersquare” (square $ABCD$). The N-1-dimensional plane $\sum_{i=1}^{N}x_{i}=d$ intersects this square along the N-2-dimensional plane $\sum_{i=1}^{N-1}x_ {i}=d$, and this last plane is perpendicular to the diagonal of the square. Therefore, the cut-off area $S_{0}=V(N-1,d)$ . Similarly, the intersection with the face $x_{N}=1$ occurs along the N-2-dimensional plane $\sum_{i=1}^{N-1}x_{i}=d-1$, and therefore $S_{ 1}=V(N-1,d-1)$. As a result, we obtain the recurrent formula
$$V(N,d)=\frac{dV(N-1,d)+(N-d)V(N-1,d-1)}{N}, N\epsilon \textbf{N}.$$
For this formula to be completely correct, it is necessary to put, by definition, $V(N,d)=0$ for $d\leq0$ , and $V(N,d)=1$ for $d\geq N.$ Using the method of mathematical induction can be proven:
$$V(N,d)=\frac{1}{N!}\sum_{k=0}^{[d]}(-1)^{k}C_{N}^{k}(d-k) ^{N}.$$
Further
$$V_{N-1}=V(N-1,\beta N)-V(N-1,\beta N-1)=\sum_{k=0}^{[\beta N]}\frac {(-1)^{k}(\beta N-k)^{N-1}N}{k!(N-k)!}.\Box$$

The exact formula \eqref{30} is very inconvenient for analysis. For such an analysis, the following theorem is needed, which gives the asymptotic value of the volume $V_{N-1}$ of the region $G'_{N}$ for $N\rightarrow\infty.$
\begin{theorem}
Let the asymptotic formula for the volume of the region $G'_{N}$ from \eqref{200} be given by the following expression:
$$V_{N-1}(\beta)=\frac{A(\beta)}{\sqrt{N}}(f(\beta))^{N}(1+o(1)),$$
where $A(\beta)$ is some positive, limited, continuously differentiable function. Then $\ A(\frac{1}{2})=\sqrt{\frac{6}{\pi}} $ and $ f(\beta)$ depends parametrically: \be f(z)=\frac{z^{\frac{z}{1-z }}(z-1)e}{\ln z},\ \beta(z)=\frac{1}{1-z}+\frac{1}{\ln z},\label{333} \ee
$$z\epsilon(0;1)\bigcup(1;+\infty),\ \beta \epsilon[0;1],\  \beta(0)=1,\  \beta(1)=\frac{1}{2},\  \beta(+\infty) =0,\  f(\beta=0)=f(\beta=1)=0,$$  $$f(\beta=\frac{1}{2})=1.$$
\end{theorem}
{\bf Proof.} The volume of $V_{N-1}$ at $N\rightarrow\infty$ can be considered approximately equal to the volume of an N-1-dimensional prism:
\be V_{N-1}=\frac{S(N-2, d=\beta(N-1))}{\sqrt{N-1}}(1+o(1)),\label{33}\ee
where $S(N-2, d=\beta(N-1))$ is the volume (area) of an N-2 -dimensional section of an N-1 -dimensional cube $0\leq y_{i}\leq1$ by the hyperplane $\sum_ {i=1}^{N-1}y_{i}=d=\beta(N-1).$ Distance from point $O(\beta,\beta,...,\beta)$ to N- 3-dimensional section of an N-2-dimensional "hypersquare"
  $y_{i}=0$ of hyperplane $\sum_{i=1}^{N-1}y_{i}=d=\beta(N-1)$ is equal to $\frac{d}{(N-1 )\cos \alpha},$ where $\alpha$ is the angle between the diagonal and the face of an N-1-dimensional cube. Obviously $\cos \alpha =\sqrt{\frac{N-2}{N-1}}.$ Next, the distance from the point $O(\beta,\beta,...,\beta)$ to N-3 -dimensional section N-2 - dimensional "hypersquare" $y_{i}=1$ of hyperplane $\sum_{i=1}^{N-1}y_{i}=d=\beta(N-1)$ is equal to $\frac{N-1-d}{(N-1)\cos \alpha}.$ Applying reasoning similar to that used in the proof of the previous theorem, we obtain the following recurrent formula:
\be S(N-2,d)=\frac{\sqrt{N-1}(dS(N-3,d)+(N-1-d)S(N-3,d-1))} {(N-2)\sqrt{N-2}},\ 0\leq d\leq N-1,\label{32}\ee
$N\geq 3,\ S(N-2,d)=0,\ d\leq 0$ or $d\geq N-1,\ N\geq 2,\ S(0,d)=1,\ 0<d<1.$

We will look for $S(N-2,d)$ in the following form:
\be S(N-2,d)=A(\frac{d}{N-1})(f(\frac{d}{N-1}))^{N}(1+o(1) ),N\rightarrow \infty .\label{31}\ee
Substituting \eqref{31} into \eqref{32}, we get:
$$(f(\frac{d}{N-1}))^{N}=(\beta(f(\frac{d}{N-2}))^{N-1}+(1- \beta)(f(\frac{d-1}{N-2}))^{N-1})(1+o(1)).$$
Further
$$(f(\frac{d}{N-1}))^{N}=(\beta(f(\frac{d}{N-1})+f'(\frac{d}{N -1})\frac{d}{(N-1)(N-2)})^{N-1}+$$
$$+(1-\beta)(f(\frac{d}{N-1})+f'(\frac{d}{N-1})\frac{d-N+1}{(N -1)(N-2)})^{N-1})(1+o(1)).$$
Let's divide both sides of the last equation by $(f(\frac{d}{N-1}))^{N-1},$ taking into account that $\frac{d}{N-1}=\beta:$
$$f(\beta)=(\beta(1+\frac{\beta f'(\beta)}{f(\beta)(N-2)})^{N-1}+
(1-\beta)(1+\frac{(\beta-1) f'(\beta)}{f(\beta)(N-2)})^{N-1})(1+o( 1)).$$
Letting N goes to infinity, we obtain the following differential equation for $f(\beta)$:
$$f=\beta e^{\frac{\beta f'}{f}}+(1-\beta) e^{\frac{(\beta-1)f'}{f}}.$$
To solve this equation, we make the following substitutions:
$$e^{\frac{f'}{f}}=z, \beta=\frac{1}{1-z-\frac{1}{V(z)}}.$$
As a result, we obtain a linear equation:
$$V'(z)=\frac{2V(z)}{1-z}+\frac{1}{z(1-z)}.$$
Choosing the integration constant from the condition:
$$ f(\beta=\frac{1}{2})=1,$$
we obtain for $f(\beta)$ the expression and all the conditions specified in the formulation of Theorem 2.3. To obtain the value of $A(\beta=\frac{1}{2})$, note that
$$\sum_{k=1}^{N-1}V_{N-1}(\beta=\frac{k}{N})=1.$$
Using \eqref{33} and \eqref{31}, we get:
\be\frac{1}{\sqrt{N-1}}\sum_{k=1}^{N-1}A(\frac{k}{N})(f(\frac{k}{N }))^{N}=1+o(1).\label{34}\ee

Consider the following integral:
$$I=\int_{0}^{1}A(\beta)(f(\beta))^{N}d\beta=\int_{0}^{+\infty}A(\beta(z ))(f(\beta(z)))^{N}\beta'(z)dz$$
Applying Laplace's method for this integral, we have:
$$I=A(\beta=\frac{1}{2})\sqrt{\frac{\pi}{6N}}(1+o(1)), N\rightarrow\infty.$$
Further
$$|I-\frac{1}{N}\sum_{k=1}^{N-1}A(\frac{k}{N})(f(\frac{k}{N})) ^{N}|\leq \frac{2}{N}.$$
Since $I=O(\frac{1}{\sqrt{N}})$, then
$$\lim_{N\rightarrow\infty}\frac{\frac{1}{N}\sum_{k=1}^{N-1}A(\frac{k}{N})(f(\frac{k}{N}))^{N}}{I}=1.$$
Then from \eqref{34} we get:
$$ A(\beta=\frac{1}{2})=\sqrt{\frac{6}{\pi}}.
$$
Now let's prove that the function \eqref{333} is single-valued. To do this, let's make a replacement:
$$\beta(u)=\frac{1-u}{2},\ z=e^{-2y}. $$
We get:
\be u=\frac{1}{y}-\coth(y).
\ee
Further
\be \beta'_{z}=\frac{\beta'_{u}u'_{y}} {z'_{y}} =\frac{e^{y}}{4}(\frac{1} {sinh^{2}y}-\frac{1}{y^{2}})\leq0,
\ee
   which proves the uniqueness of the function \eqref{333}.$\Box$

  The polyhedra method gave us the opportunity to obtain an asymptotic formula for estimating $C_{N}$, but obtain the value of the limit $b_{N}$ using this method
  failed. The following method provides this opportunity.
\section{Recurrent convolution method}
Let's change our model a little. Let's remove from \eqref{25} all the constants by which the integrand exponent is multiplied and make the following substitutions:
\be u_{i}=1-2y_{i},\ i=1,...,N,\ u=1-2\beta,\ mc^{2}(t_{2}-t_{1} )=t. \ee
Let us normalize the integrand by $1$ and write the resulting functional sequence in a slightly different form:
\begin{eqnarray}
p^{(N)}(t,u)= \int\delta(u-\frac{1}{N}\sum_{i=1}^{N}
u_i)\prod_{i=1}^{N}P(\frac{t}{N},u_i)du_i\label{100}
,\end{eqnarray}
where
\begin{eqnarray}
\label{1} P(t,u)=\frac{\exp \left(t \sqrt{1-{u}^{2}}\right)\left(\Theta(u+1)- \Theta (u-1)\right)}{\int_{-1}^{1}\exp \left(t \sqrt{1-{u}^{2}}\right)du}.
\end{eqnarray}
  Here the integration is carried out over $N-1$ - a dimensional section of the N-dimensional cube $|u|\leq1$, which passes through the point $(u,\ u, ...,u)$ perpendicular to the diagonal connecting the points $( -1,\ -1,...,-1)$ and $(1,\ 1,...,1).$ We will examine the resulting sequence for asymptotics. Note that this sequence can also be written in recurrent form:
\be p^{(N)}(t,u)=p^{(N)}(N\tau,u)=\frac{\int_{-1}^{1}p^{N-1} ( (N-1)\tau,\ \frac{Nu-u_{N}}{N-1})e^{\tau\sqrt{1-u_{N}^{2}}}du_{N} }{\int_{-1}^{1} e^{\tau\sqrt{1-u^{2}}}du}. \label{35}
\ee
For the denominator we have:
\be \int_{-1}^{1} e^{\tau\sqrt{1-u^{2}}}du=2+\frac{\pi \tau}{2}+O(\frac{ 1}{N^{2}}).
\ee
We will look for asymptotics in the form:
\be p^{(N)}(N\tau,u)=A_{N}(y(u))e^{N f(\tau,y(u))}(1+O(\frac{ 1}{N})). \label{600}
\ee
The functions $f(\tau,y(u)),\ y(u)$ are to be defined.
For the purposes of this chapter, the exact form of $A_{N}(y(u))$ does not matter, but only its following property matters:
\be \lim_{N\rightarrow \infty}\frac{A_{N}(y(u))}{A_{N-1}(y(u))}=1.
\ee
We will look for $f(\tau,y(u))$ among bounded functions that are twice continuously differentiable on R with respect to all variable. In addition, the required function must be even in $u$ and turn to $0$ when $|u|\geq 1$. Using Taylor's formula we get:
\be f(\tau,y(u))=f(0,y(u))+\tau f_{\tau}(0,y(u))+O(\frac{1}{N^{ 2}}).
\ee
Let's introduce the following notation:
\be f(0,y(u))=f_{1}(u),\ f_{\tau}(0,y(u))=f_{2}(u) \label{500}
\ee.
\begin{theorem}
Let the asymptotic formula for $p^{(N)}(N\tau,u)$ have the form \eqref{600}-\eqref{500}, then
the function $f_{1}(u)$ has a parametric dependence:
\be f_{1}(y)=uy+\ln(\frac{\sinh(y)}{y}),\
u=\frac{1}{y}-\coth(y),
\ee
$$y\epsilon(-\infty;+\infty),\ u \epsilon[-1;1],\  u(-\infty)=1,\  \ u(0)=0,\  u(+\infty) =-1,\ $$
$$f_{1}(u=-1)=f_{1}(u=1)=-\infty,\ f_{1}(0)=0.$$
\end{theorem}
{\bf Proof.} Using all the above facts and notations, from \eqref{35} for $\tau=0$ we have:
\be e^{Nf_{1}(u)}(1+o(1))=\frac{1}{2}\int_{-1}^{1}e^{(N-1)f_{1}(u+\frac{u -u_{N}}{N-1})}du_{N}.
\ee
Further
\be e^{Nf_{1}(u)}(1+o(1))=\frac{1}{2}\int_{-1}^{1}e^{(N-1)f_{1}(u)+f_{ 1}^{\prime}(u)(u-u_{N})}du_{N},\label{300}
\ee
For the leading order of the asymptotic equality \eqref{300} we have:
\be e^{f_{1}(u)}=\frac{1}{2}\int_{-1}^{1}e^{f_{1}^{\prime}(u)(u- u_{N})}du_{N},
\ee
\be e^{f_{1}(u)}=\frac{e^{f_{1}^{\prime}(u)(u+1)}-e^{f_{1}^{\prime }(u)(u-1)}}{2f_{1}^{\prime}(u)},
\ee
\be e^{ f_{1}(u)}=\frac{e^{uf_{1}^{\prime}(u)}\sinh(f_{1}^{\prime}(u))} {f_{1}^{\prime}(u)},
\ee
\be f_{1}(u)=u f_{1}^{\prime}(u)+\ln(\frac{\sinh(f_{1}^{\prime}(u))}{f_{ 1}^{\prime}(u)}).
\ee
As a result, we obtained the famous Clairaut equation. We solve it in a known way. We introduce the parameter $f_{1}^{\prime}(u)=y$ and differentiate the last equation. Unphysical solutions $f_{1}(u)=Cu+\ln(\frac{\sinh(C)}{C})$ ($C$ is an arbitrary constant) are discarded. There remains a special solution:
\be f_{1}(y)=uy+\ln(\frac{\sinh(y)}{y}),\
u=\frac{1}{y}-\coth(y). \label{400}
\ee
Further
\be u'(y)=\frac{1} {sinh^{2}y}-\frac{1}{y^{2}}\leq0,
\ee
   which proves the uniqueness of the function \eqref{400}.$\Box$

Considering that
\be p^{(N)}(0,u)=\frac{N}{2}V_{N-1}(\beta(u)),\ \beta(u)=\frac{1-u}{2}\label{101}
\ee
and making the change $z=e^{-2y}$, we obtain a complete coincidence of the results of Theorems 3.1 and 2.3.
\begin{theorem}
Let the asymptotic formula for $p^{(N)}(N\tau,u)$ have the form \eqref{600}-\eqref{500}, then
the function $f_{2}(u)$ has a parametric dependence:
\be f_{2}(y)=\frac{\pi}{2}(\frac{I_{1}(y)}{\sinh y}-\frac{1}{2}),\ u= \frac{1}{y}-\coth(y),
\ee
where $y\epsilon(-\infty;+\infty),\ f_{2}(0)=0,$ $I_{1}(y)$ is the modified first-order Bessel function.
\end{theorem}
{\bf Proof.}From \eqref{35} after eliminating infinitesimals of higher order than $\frac{1}{N}$, we obtain:
\be e^{N(f_{1}(u)+\tau f_{2}(u))}=\frac{\frac{1}{2}\int_{-1}^{1}e^ {(N-1)(f_{1}(u)+\tau f_{2}(u))+(f_{1}^{\prime}(u)+\tau f_{2}^{\prime }(u))(u-u_{N})+\tau\sqrt{1-u_{N}^{2}}}du_{N}}{1+\frac{\pi\tau}{4} }.
\ee
Let us reduce the resulting expression by $e^{(N-1)(f_{1}(u)+\tau f_{2}(u))}$, then take the derivative with respect to $\tau$ from both sides of this expression and let's equate $\tau$ to $0$:
\be e^{f_{1}}f_{2}=\frac{1}{2}\left(\int_{-1}^{1}e^{f_{1}^{\prime}(u )(u-u_{N})}f_{2}^{\prime}(u)(u-u_{N})du_{N}+\int_{-1}^{1}
e^{f_{1}^{\prime}(u)(u-u_{N})}(\sqrt{1-u_{N}^{2}}-\frac{\pi}{4}) du_{N}\right).
\ee
The first integral on the right side of the last expression turns out to be equal to $0$ after calculations. After calculating the second integral we get:
\be f_{2}(y)=\frac{\pi}{2}(\frac{I_{1}(y)}{\sinh y}-\frac{1}{2}),\ u= \frac{1}{y}-\coth(y).\Box
\ee
The recurrent convolution method gives us the time dependence of the first term of the asymptotics accurate to a pre-exponential factor. The value of this factor and the ability to obtain other terms of the asymptotics gives us the following method.
\section{Saddle point method}
Let's put $t=0$ in \eqref{100} and perform direct and inverse Fourier transforms:
\begin{eqnarray}
\label{17} p^{(N)}(0,u)=\frac{1}{2\pi}\int_{-\infty}^{+\infty}(\frac{\sin\frac{ \omega}{N}}{\frac{\omega}{N}})^{N}e^{-i\omega u}d\omega. \end{eqnarray}
Using the substitution $\frac{\omega}{N}=\varpi$ and formula (11) on page 28 of \cite{BE}, we obtain the exact value of this integral:
\begin{eqnarray}
\label{20} p^{(N)}(0,u)=\frac{ N^{2}}{2^{N}}\sum_{k=0}^{[\frac{(1+ u)N}{2}]}\frac{(-1)^{k}((1+u)N-2k)^{N-1}}{k!(N-k)!}.
\end{eqnarray}
Given \eqref{101}, we have an exact match with \eqref{30} from Theorem 2.2.
In addition, \eqref{20} was also used in \cite{Sen} to improve the accuracy of approximations
in the central limit theorem. Although this formula is accurate, it is impossible
obtain from it the asymptotics for \eqref{17}. To obtain this asymptotics, we use the saddle point method. After replacement we get:
\begin{eqnarray}
p^{(N)}(0,u)=\frac{N}{2\pi}\int_{-\infty}^{+\infty}\exp(N(\ln(\frac{\sin\varpi}{\varpi})-i\varpi u)) d\varpi.\label{6}
\end{eqnarray}
Let us now prove the following lemma.
\begin{lemma}
 Let the integral of a complex-valued function of a real variable converge absolutely $\int_{\mathbb{R}}
|g(\omega)|d\omega<\infty$  and the function module is limited to the integration area $|g| \leq g_0 < 1$, then the integral of its degree  $ \int_{\mathbb{R}} g^N
d\omega\rightarrow 0$ has the order of smallness $O(e^{-N |\ln g_0|})$   at $N\rightarrow\infty$.
\end{lemma}
{\bf Proof.}
The proof obviously follows from next chain of inequalities:\\
$$|\int g^N d\omega|\leqslant \int |g|^Nd\omega
\leqslant g_0^N \int |g/g_0|^Ndu\leqslant
g_0^N \int |g/g_0|du \leqslant g_0^N ,$$ \\
therefore $ |\int g^N d\omega| = O(e^{-N |\ln g_0|}).\Box$

\begin{theorem}
The asymptotic formula for $p^{(N)}(0,u)$ from \eqref{17} is given by the following expression:
  $$p^{(N)}(0,u)=y \sinh y \sqrt{\frac{N}{2\pi(\sinh^{2} y-y^{2})}}(\frac{ \sinh y}{y}e^{uy})^{N}(1+O(\frac{1}{{N}})),$$
  \begin{eqnarray}u=\frac{1}{y}-\coth(y),\label{102}
\end{eqnarray}
where $y\epsilon(-\infty;+\infty).$
\end{theorem}
{\bf Proof.} We present formula (18) on page 452, 453 from \cite{Lav_Sh}:
\begin{eqnarray}
\label{9} \int_{C}\varphi(\varpi)e^{Nf(\varpi)}d\varpi=e^{Nf(\varpi_{0})}
\sqrt{\frac{2\pi}{N|f''(\varpi_{0})|}}\varphi(\varpi_{0})(1+O(1/N)), N\rightarrow \infty.
\end{eqnarray}
In our case, contour C is the real axis. The integrand is analytical, and therefore the contour can be deformed as desired. Next, $\varphi(\varpi)\equiv 1$, $f(\varpi)=\ln(\frac{\sin\varpi}{\varpi})-i\varpi u$, $\varpi_{0} $ -\ the saddle point, i.e., the solution to the equation $f'(\varpi)=0$. This equation has the following form:
\begin{eqnarray}\label{3} \cot \varpi-\frac{1}{\varpi}=iu.
\end{eqnarray}
Equating the real and imaginary parts in \eqref{3} respectively, we obtain:
\begin{eqnarray}
  \frac{x}{x^{2}+y^{2}}=\frac{\sin2x}{\cosh2y-\cos2x}, \label{4}
\end{eqnarray}
\begin{eqnarray}
\label{12} u=\frac{sh2y}{\cos2x-\cosh2y}+\frac{y}{x^{2}+y^{2}},
\end{eqnarray}
where $\varpi=x+iy$. One solution \eqref{4} is obvious: $x=0$. The solution in this case will be purely imaginary:

\begin{eqnarray}
  \varpi_{0}=iy, u=\frac{1}{y}-\coth y, \label{8}
\end{eqnarray}
where $y$ is a real number. The function $u=u(y)$ is a real function of a real variable. This function is strictly decreasing, odd, $u(-\infty)=1,\ u(0)=0, u(+\infty)=-1$. In addition, \eqref{4} has an infinite number of solutions for each fixed $u$. However, we will show below that in the saddle point method, for each fixed $u$, it is sufficient to take into account the contribution of only one critical point \eqref{8}. The formula \eqref{9} taking into account the contribution of only the point \eqref{8} gives us the asymptotics \eqref{6} in the form \eqref{102}.

Let's deform the contour of the integral \eqref{6}. Now it will pass through the main point of the pass $\varpi=iy$ along the line of the largest slope in a small neighborhood of this point. This line intersects a given point, having a horizontal tangent. The integrand in \eqref{6} on this line will be real and positive, $y$ depends on $u$ according to the formula \eqref{8}. The required integral in a small neighborhood of the main saddle point along the line of the largest slope is estimated by the Laplace method, this estimate is given by the formula \eqref{102}. For $u=0$, by taking the limit at $y\rightarrow 0$, this estimate becomes the following:
\begin{eqnarray}p^{(N)}(0,0)=\sqrt{\frac{3N}{2\pi}}
(1+ O(\frac{1}{N})),\ N\rightarrow\infty.
\end{eqnarray}
Beyond the small neighborhood of the main saddle point, we continue the integration contour with two horizontal rays. Obviously, the integral along such a new contour will be equal to the integral along the old one, i.e., along the real axis. Let us prove that the integral over such horizontal rays is exponentially small compared to the estimate \eqref{102}. For $u=0$ and any positive $\varepsilon$ it is necessary to evaluate the following integral:
\begin{eqnarray} 2 \frac{N}{2\pi}\int_{\varepsilon}^{+\infty}(\frac{\sin x}{x})^{N} d x= \frac{N }{\pi}\int_{\varepsilon}^{+\infty}((\frac{\sin x}{x})^{3})^{\frac{N}{3}} d x. \label{39}
\end{eqnarray}
Integral
\begin{eqnarray}
\int_{\varepsilon}^{+\infty}(\frac{\sin x}{x})^{3} d x
\end{eqnarray}
obviously absolutely converges. For a sufficiently small and positive $\varepsilon$, the integrand is always strictly less than 1 in modulus:
\begin{eqnarray}
|(\frac{\sin x}{x})^{3}|\leq (\frac{\sin \varepsilon}{\varepsilon})^{3}< 1.
\end{eqnarray}
Therefore, by Lemma 4.1, the integral \eqref{39} exponentially tends to 0. Now let us consider the case when $u\neq 0$ and, consequently, $y\neq 0$.
Let the origins of the rays have coordinates $(-\varepsilon;\ y_{1})$ and $(\varepsilon;\ y_{2})$. In this case we get:
\begin{eqnarray}\nonumber
&&|\frac{N}{2\pi}\int_{-\infty}^{-\varepsilon}(\frac{\sin(x+i y_{1})}{x+i y_{1}} )^{N}e^{-iuxN+y_{1}uN} dx+
\frac{N}{2\pi}\int_{\varepsilon}^{+\infty}(\frac{\sin(x+i y_{2})}{x+i y_{2}})^{ N}e^{-iuxN+y_{2}uN} dx| \\\nonumber
&\leq&\frac{N}{2\pi}\int_{\varepsilon}^{+\infty}(\frac{\sin^{2}(x)+\sinh^{2}(y_{1} )}{x^{2}+ y_{1}^{2}})^{\frac{N}{2}}e^{y_{1}uN} dx+
\frac{N}{2\pi}\int_{\varepsilon}^{+\infty}(\frac{\sin^{2}(x)+\sinh^{2}(y_{2})}{ x^{2}+ y_{2}^{2}})^{\frac{N}{2}}e^{y_{2}uN} dx \nonumber
\\\label{130}
&\leq&\frac{N}{\pi}\int_{\varepsilon}^{+\infty}(\frac{\sin^{2}(x)+\sinh^{2}(y_{1}) }{x^{2}+ y_{1}^{2}})^{\frac{N}{2}}e^{y_{1}uN} dx.
\end{eqnarray}
The last inequality is true if the first integral in \eqref{130} is greater than or equal to the second one. If this is not the case, then the further proof is similar. Let us now prove that for small positive $\varepsilon$ and for any $x>\varepsilon$
\begin{eqnarray}
\frac{\sin^{2}(x)+\sinh^{2}(y_{1})}{x^{2}+ y_{1}^{2}}<\frac{\sin^{ 2}(\varepsilon)+\sinh^{2}(y_{1})}{\varepsilon^{2}+ y_{1}^{2}}.
\end{eqnarray}
Really
\begin{eqnarray}
\frac{\sin^{2}(\varepsilon)+\sinh^{2}(y_{1})}{\varepsilon^{2}+ y_{1}^{2}}-\frac{\sin ^{2}(x)+\sinh^{2}(y_{1})}{x^{2}+ y_{1}^{2}}=
\end{eqnarray}
\begin{eqnarray}
=\frac{x^{2}\varepsilon^{2}(\frac{\sin^{2}(\varepsilon)}{\varepsilon^{2}}-\frac{\sin^{2}(x )}{x^{2}})+\sinh^{2}(y_{1})[(x^{2}-
\frac{y_{1}^{2}}{\sinh^{2}(y_{1})}\sin^{2}(x))-(\varepsilon^{2}-\frac{y_{ 1}^{2}}{\sinh^{2}(y_{1})}\sin^{2}(\varepsilon))]
}{(\varepsilon^{2}+ y_{1}^{2})(x^{2}+ y_{1}^{2})}.
\end{eqnarray}
The resulting expression is positive, since the function $y=\frac{\sin^{2}(x)}{x^{2}}$ at the point $x=0$ has an absolute maximum, and the function $y= x^ {2}-k\sin^{2}(x)$ is increasing for any $x>0$ and $0<k<1$.
Further
\begin{eqnarray}
\frac{\sin^{2}(\varepsilon)+\sinh^{2}(y_{1})}{\varepsilon^{2}+ y_{1}^{2}}e^{2y_{1 }u}<\frac{\sinh^{2}(y)}{y^{2}}e^{2uy}.
\end{eqnarray}
The last inequality is obtained due to the fact that the deformed contour of the integral \eqref{6} in the vicinity of the main saddle point goes along the line of the largest slope. Then we have:
\begin{eqnarray}
  \frac{N}{\pi}\int_{\varepsilon}^{+\infty}\left(\frac{\sin^{2}(x)+\sinh^{2}(y_{1})} {x^{2}+ y_{1}^{2}}\right)^{\frac{N}{2}}e^{y_{1}uN} dx=
\end{eqnarray}
\begin{eqnarray}
=\frac{N}{\pi}(\frac{\sinh y}{y}e^{uy})^{N}\int_{\varepsilon}^{+\infty}\left(\frac{\frac{\sin^{2}(x)+\sinh^{2}(y_{1})}{x^{2}+ y_{1}^{2}}e^{2y_{1}u} }{\frac{\sinh^{2} y}{y^{2}}e^{2uy}}\right)^{\frac{N}{2}}dx. \label{40}
\end{eqnarray}
Integral
\begin{eqnarray}
\int_{\varepsilon}^{+\infty}\frac{\frac{\sin^{2}(x)+\sinh^{2}(y_{1})}{x^{2}+ y_{ 1}^{2}}e^{2y_{1}u}}{\frac{\sinh^{2} y}{y^{2}}e^{2uy}}dx
\end{eqnarray}
obviously absolutely converges. The integrand in it, for a sufficiently small and positive $\varepsilon$, according to the modulus proved above, is always strictly less than 1. Therefore, by Lemma 4.1, the integral from \eqref{40} exponentially tends to 0.
Thus, the theorem is proven.$\Box$

The result obtained in Theorem 4.1 coincides exactly with the results of Theorems 2.3 and 3.1. The resulting asymptotics tells us that $p^{(N)}(0,u)$ tends to the delta function as $N$ tends to infinity: when $u=0$ the sequence tends to infinity as $\sqrt {N}$, and for the remaining $u$ exponentially tends to zero. In this case, the integral of the sequence remains equal to 1.

Now let's study the asymptotics of $p^{(N)}(t,u)$.
\begin{theorem}
The asymptotic formula for $p^{(N)}(t,u)$ from \eqref{100} is given by the following expression:
\begin{eqnarray}p^{(N)}(t,u)=p^{(N)}(0,u)\exp(\frac{\pi t}{2}(\frac{I_{1 }(y(u))}{\sinh y(u)}-\frac{1}{2}))(1+O(\frac{1}{N})),\label{110}
\end{eqnarray}
where  $y(u)$ is defined by next equation :
\begin{eqnarray}
u=\frac{1}{y}-\coth y, \label{110+}
\end{eqnarray}
where  $t\geq 0,$ $y\epsilon(-\infty;+\infty)$ and $I_{1}(y)$ is the modified first-order Bessel function.
\end{theorem}
{\bf Proof.}
The exact value of $p^{(N)}(t,u)$, obtained from \eqref{100} by direct and inverse Fourier transform, has the form:
\begin{eqnarray}p^{(N)}(t,u)=\frac{1}{2\pi}\int_{-\infty}^{+\infty}\left(\frac{\int_{ -1}^{1}e^{\frac{1}{N}
(i\omega u+t\sqrt{1-u^{2}})}du}{\int_{-1}^{1}e^{\frac{t}{N}
\sqrt{1-u^{2}}}du}\right)^{N}e^{-i\omega u}d\omega.
\end{eqnarray}
Further
\begin{eqnarray}p^{(N)}(t,u)=\frac{1}{2\pi}\int_{-\infty}^{+\infty}\left(\frac{\int_{ -1}^{1}e^{\frac{1}{N}
i\omega u}(1+\frac{t\sqrt{1-u^{2}}}{N}+O(\frac{1}{N^{2}}))du}{2+\frac{\pi t}{2N}+
O(\frac{1}{N^{2}})}\right)^{N}e^{-i\omega u}d\omega.
\end{eqnarray}
Using the substitution $\frac{\omega}{N}=\varpi$ and formula (9.1.20) on page 182 of \cite{Abr_St}, we get:
\begin{eqnarray}p^{(N)}(t,u)=\frac{N}{2\pi}\int_{-\infty}^{+\infty}
\left(\frac{\sin\varpi}{\varpi}+\frac{\pi t J_{1}(\varpi)}{2N\varpi}\right)^{N}e^{-i\varpi uN-\frac{\pi t}{4}} d\varpi(1+O(\frac{1}{N})).
\end{eqnarray}
Here $J_{1}$ is a first-order cylindrical Bessel function. Further
\begin{eqnarray}p^{(N)}(t,u)=\frac{N}{2\pi}\int_{-\infty}^{+\infty}
(\frac{\sin\varpi}{\varpi})^{N}e^{-i\varpi uN-\frac{\pi t}{4}} (1+\frac{\pi t J_{1 }(\varpi)}{2N\sin\varpi})^{N}d\varpi(1+O(\frac{1}{N})),
\end{eqnarray}
\begin{eqnarray}
\label{13} p^{(N)}(t,u)=\frac{N}{2\pi}\int_{-\infty}^{+\infty}
(\frac{\sin\varpi}{\varpi})^{N}e^{-i\varpi uN} \exp(\frac{\pi t}{2}(\frac{J_{1}(\varpi)}{\sin\varpi}-\frac{1}{2}))d\varpi(1+O(\frac{1}{N})).
\end{eqnarray}

Further, repeating all the reasoning of the previous theorem, we obtain \eqref{110}. The theorem is proven. $\Box$
\section{Remarks on a possible approach to relativistic Brownian motion}
 The result obtained in Theorem 4.2 coincides exactly with the result of Theorem 3.2. The sequence $p^{(N)}(t,u)$ for $t>0$ tends to the delta function in the same way as for $t=0$. The saddle point method can also be used to obtain other terms of the asymptotic series.
Next, the regularized limit (i.e. equal to a finite positive number) will be:
\begin{eqnarray}
\label{15} p(t,u)=\lim_{N\rightarrow\infty}\frac{p^{(N)}(t,u)}{p^{(N)}(0,u)}=\exp(\frac {\pi t}{2}(\frac{I_{1}(y)}{\sinh y}-\frac{1}{2})),\ u=\frac{1}{y}-\ coth y.
\end{eqnarray}
If we normalize this limit and plot it as a function of $u$ with parameter $t$, then for $t=0$ we will get a uniform distribution, and then as $t$ grows we will see that the functions will be even with a single maximum at zero and equal minima at $u=\pm 1$. As $t$ increases, the maximum will increase, tending to infinity, while the minimum will decrease, tending to zero. Such a sequence obviously tends to a delta function. In conclusion, we present a formula for calculating the probability of transition of a Brownian particle from a point with coordinate $x$ to a point lying on the segment $[x+u_{1}t;\ x+u_{2}t]$ (if  after normalization take \eqref{15} as the density of such probability):
\begin{eqnarray}
\label{16} p=\frac{\int_{y_{2}}^{y_{1}}\exp(\frac{\pi t}{2}(\frac{I_{1}(y)} {\sinh y}-\frac{1}{2}))(\frac{1}{y^{2}}-\frac{1}{\sinh^{2}y})dy}{\int_ {-\infty}^{+\infty}\exp(\frac{\pi t}{2}(\frac{I_{1}(y)}{\sinh y}-\frac{1}{2} ))(\frac{1}{y^{2}}-\frac{1}{\sinh^{2}y})dy},
\end{eqnarray}
  where $y_{k}$ is the root of the equation $u_{k}=\frac{1}{y_{k}}-\coth y_{k},\ k=1,2.$

In \eqref{15} $p^{(N)}(0,u)$ has the meaning of the cross-sectional area of the cube over which the integration is carried out. Thus, the regularized limit has the meaning of the average value of the exponent of the minus action over all trajectories. This average is taken at a given value of the average speed $u$. If we expand \eqref{1} using the Taylor formula and apply the mean value theorem, then for small $u$ we obtain:
\begin{eqnarray}
P(t,u)=\frac{1}{2}exp(-\frac{tu^{2}}{2}+t(1-\sqrt{1-u^{2}_{1}}))(1+O(u)),\
0<u_{1}<1.
\end{eqnarray}
If we do the same with normalized \eqref{15}, we get:
\begin{eqnarray}
p_{nor}(t,u)=\frac{1}{2}exp(-\frac{3\pi tu^{2}}{8}-\frac{\pi t}{2}(\frac{I_{1}(y_{0})}{\sinh y_{0}}-\frac{1}{2}))(1+O(u)).
\end{eqnarray}
As you can see, the distribution at both the beginning and the end is almost Gaussian, but with different values of the standard deviation.\\

It is of interest to compare the dispersion of the relativistic random walk for a regularized probability density with the dispersion of non-relativistic Brownian motion. The variance for the regularized probability density is calculated using the formula:
\begin{eqnarray}
\label{16} \sigma^{2}_{rel}(t)=\frac{\int_{-\infty}^{+\infty} \exp(\frac{\pi t}{2}(\frac{I_{1}(y)} {\sinh y}-\frac{1}{2}))(\frac{1}{y^{2}}-\frac{1}{\sinh^{2}y})(\frac{1}{y}-coth(y))^{2}dy}{\int_ {-\infty}^{+\infty}\exp(\frac{\pi t}{2}(\frac{I_{1}(y)}{\sinh y}-\frac{1}{2} ))
(\frac{1}{y^{2}}-\frac{1}{\sinh^{2}y})dy},
\end{eqnarray}
For a nonrelativistic probability density, the average velocity dispersion is: $\sigma^{2}_{nonrel}(t)= 1/t$

The variance estimation was performed numerically.
  We might expect that, due to the speed limit, the dispersion of a relativistic walk will always be lower than that of a non-relativistic walk, but this property is observed only at small times. Then the dispersion calculated from the regularized probability density becomes higher than that of non-relativistic Brownian motion, and only at very long times a new decrease in the average speed is observed. The counterintuitive behavior of the moments of the adjusted probability density makes the justification of such a regularization procedure problematic and requires further research.

The ratio of probability densities, generally speaking, may not be a probability density, because the division operation preserves only the property of non-negativity, but not the properties of integrability and normalization. In particular,  there is uncertainty beyond the support of the regularized measure. Therefore, regularization of the measure, although it is a remarkable mathematical fact, may not be the best way to build a model of relativistic Brownian motion. Non-relativistic Brownian motion does not require regularization. Therefore, the necessity to regularize the path integral in the relativistic case may mean the need to weaken the requirements for the relativistic random walk model either in terms of the compactness of the support or in terms of the markovity of the process.

\section{Conclusion}
So, we have shown that our straightforward reasoning about the random walk with
limited speed based on the path integral in
configuration space leads to a degenerate
distribution with support of measure zero. Besides, the rate of convergence and asymptotic behavior of these integrals were investigated. The study was conducted using three different methods, which yielded consistent results. Along the way, exact and asymptotic formulas were obtained for the volume of some parts and sections of an $N$-dimensional unit cube. We also discussed the possibility of using a renormalized path integral to obtain the transition probability of relativistic Brownian motion.

  All calculations were performed mainly by E. A. Kurianovich. The authorship of A. I. Mikhailov is the model \eqref{100}, Lemma 4.1 and the idea to apply the Fourier transform. I.V. Volovich  is responsible for the formulation of the problem and
general leadership.

\section{Acknowledgments}
We express our gratitude to the participants of the seminar on quantum mathematical physics of the scientific and educational center in Steklov Mathematical Institute RAS for numerous and fruitful discussions.\\
The study was prepared with the support of the Russian Science Foundation, Project No. 22-78-10171 "Transdisciplinary conceptualizations of scientific progress: problem-oriented, semantic and epistemic approaches. On the 100th anniversary of the birth of Thomas Kuhn and Imre Lakatos."

\end{document}